# Ultrahigh enhancement of light focusing through disordered media controlled by Mega-pixel modes

HyeonSeung Yu,[1,2] KyeoReh Lee,[1,2] YongKeun Park,[1,2*]

[1]Department of Physics, Korea Advanced Institute of Science and Technology, Daejeon 34141, Republic of Korea
[2]KAIST Institute of Health Science and Technology, Daejeon 34141, Republic of Korea
*Corresponding author: yk.park@kaist.ac.kr

We propose and demonstrate a system for wavefront shaping which generates optical foci through complex disordered media and achieves an enhancement factor of greater than 100,000. To exploit the 1 Megapixel capacity of a digital micro-mirror device and its fast frame rate, we developed a fast and efficient method to handle the heavy matrix algebra computation involved in optimizing the focus. We achieved an average enhancement factor of 101,391 within an optimization time of 73 minutes. This unprecedented enhancement factor may open new possibilities for realistic image projection and the efficient delivery of energy through scattering media.

## 1. Introduction

The ability to concentrate light through scattering media has been well demonstrated, and in recent years has generated considerable research interest [1-3]. Light impinging into a scattering media is scattered multiple times, and the resulting transmitted light field exhibits complex speckle patterns. The deterministic nature of elastic light scattering ensures that there is a linear relationship between the input and the output fields. Thus, by properly modulating the incident light field with a wavefront shaper, such as a spatial light modulator (SLM) or a digital micro-mirror device (DMD), the random speckle can be manipulated into desired patterns such as focus spots [4] or complex patterns [5, 6]. This wavefront shaping technique has been utilized to investigate a range of physical optics including spectral control [7], polarization modulation [8], spatiotemporal modulation [9], near-field control [10], active plasmonics [11], nonlinear focusing [12], enhanced Raman [13], and optical phase conjugation [14]. Furthermore, this phenomenon has also been employed in numerous applications, including imaging through biological tissues [15] or multimode fibers [16], optogenetics [17], optical coherence tomography [18], scattering optical elements [19], photovoltaics [20], and 3D holographic displays [21].

To efficiently optimize light focusing, several algorithms have been developed for SLMs, utilizing phase-only [19, 22-27] or binary amplitude control [28, 29]. In both cases, to quantify the degree of control achieved in the wavefront shaping, a value known as enhancement factor (EF) is used. EF is defined as the intensity of the optimized focus over the averaged background intensity. EF is linearly proportional to the number of independently controlled pixels in an SLM [1], and thus increasing the modes results in an increase in EF.

Because most types of wavefront shapers, including the LCoS (Liquid Crystal on Silicon) SLM and digital micro-mirror device (DMD), have several megapixels, the EF can exceed 100,000 in principle. Nevertheless, no EF above 1,100 has yet been reported, mainly due to slow optimization speeds and the lack of an efficient optimization system.

This small EF hinders wide applications of the wavefront shaping technique, including for example, image projection [5], 3-D light patterning [30] or lithography applications [31]. Such applications require considerably higher EFs, since the enhanced energy is distributed among the focused spots [1]. Furthermore, it should be noted that the energy transmission enhancement provided by small EFs is only useful for a single output mode at the microscopic scale. Let us suppose a 1 µm² sized focus is formed inside a 32 x 32 µm² square. When the EF of the single focus is given as 1,000, the total intensity inside the square is only enhanced by a factor of two. Therefore, if significant energy enhancement over a large area is required, it is also crucial to increase the number of controlling modes.

Here, we demonstrate a novel wavefront shaping system which provides ultrahigh EFs by exploiting Mega-mode control of a DMD. Based on measurements of the transmission matrix of the scattering media, the optimization pattern is calculated using a modified version of the Hadamard basis algorithm [29] and parallelized computation. In an optimization time of 73 min, we achieved an EF of 101,391 on average, which is approximately 100 times higher than the best record in the literature. We also discuss the advanced nature of the proposed method by comparison with existing wavefront shaping methods.

## 2. Principles

In previous studies, LCoS SLMs or DMDs have mainly been used for wavefront shaping through turbid media. Deformable mirrors (DM) also provide phase modulation at fast frame rates. However, the total number of pixels in the DM is limited to thousands [14, 32, 33]. While commercially available LCoS SLMs and DMDs provide 0.4 to 12.0 megapixels, only a small portion of the pixels were used in previous experiments. For example, the number of effective modes is reduced to hundreds or thousands by grouping adjacent pixels into macro pixels, as shown in Fig. 1(a). As a result, the best EF achieved in previous studies was 1,080 [1].

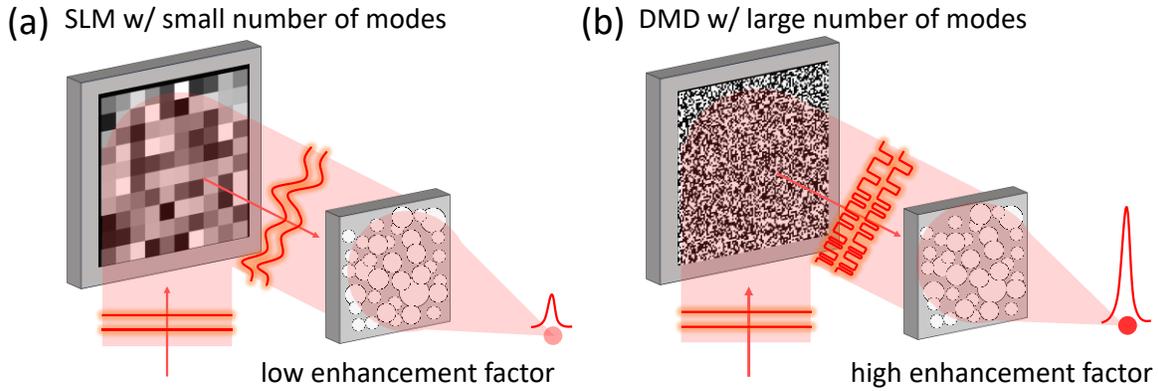

Fig. 1. (a) When SLMs have been used for focusing through turbid media, only a small number of modes, up to 3228 (super-pixels of SLMs) [1], were actually used. Thus, the optical foci that are generated have a low enhancement factor. (b) In the present work, we achieved an enhancement factor of more than 100,000 by individually controlling up to 1 megapixel with a DMD.

LCoS SLM provides a large number of pixels, but the slow refresh rate (up to 60 Hz) creates many limitations to its application. For example, generating focus with 1 million modes of the SLM requires a minimum of 19 hours using the 4-point phase shift method [1]. On the other hand, a DMD has a much faster refresh rate up to 23 kHz, and the number of controllable modes per unit time can be greatly increased. Therefore, DMD has been used for focusing through scattering mediums which exhibit rapid decorrelation, such as biological tissue [27, 34]. However, increasing the number of modes to the maximum capacity remains a very difficult task due to various technical problems, such as the large required amounts of data processing and overload calculation.

In this study, we utilized the high pixel resolution of a DMD to increase the number of control modes to 1 Megapixel. We achieved control of the light passing through a scattering layer at high speed with 1,048,576 modes, corresponding to approximately half of the total DMD of 1,920 × 1,080 pixels.

### 3. Optical Setup

The experimental setup is depicted in Fig. 2. A coherent laser beam from a He-Ne laser ($\lambda$ = 632.8 nm; 10 mW; HNL100L-EC, Thorlabs Inc., United States) was coupled into a single mode fiber (SMF) (SM600; Thorlabs Inc., United States). A neutral-density filter is used when measuring the EF. The diverging beam from the fiber output is collimated by a lens and spatially modulated by a DMD (1,920×1,080 pixels, DLP6500, Texas Instruments Inc., United States). The diffracted beams from the DMD are transmitted through two holographic diffusers (15° diffusing angle, #54-495, Edmund Optics, United States). After transmission, the output field exhibits speckle patterns with high contrast due to the multiple scattering.

For point optimization, after the second diffuser HD2, one end facet of an SMF was placed at the target focus spot. Optimizing the focus maximizes the amount of light that is coupled to the SMF. The collected light intensity is measured at the other end of the fiber using a single photon counting module (SPCM) (SPCM-AQRH-11, Excelitas Technologies Corp., United States) via a 2-$f$ imaging system. A laser line filter (LL01-633-25, Semrock Inc., United States) was inserted before the SPCM to block unwanted ambient light.

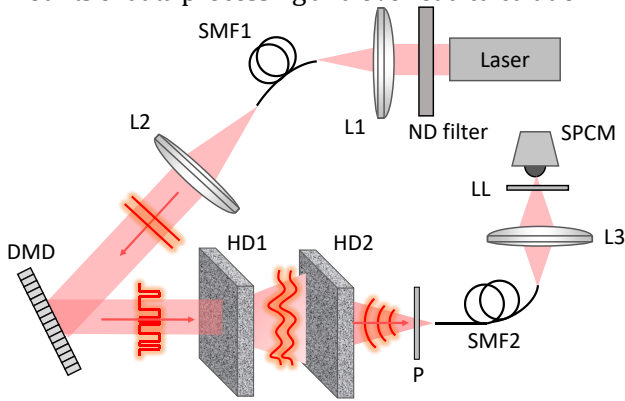

Fig. 2. Optical system for the ultrahigh enhancement of focusing. L1-3 are lenses with focal lengths of 30 mm, 50 mm, 30 mm lenses, respectively; SMF: single mode fiber; DMD: a digital micro-mirror device; HD: a holographic diffuser; P: linear polarizer; LL: laser line filter; SPCM: single photon counting module.

### 4. System Design

To achieve ultrahigh enhancement of the light focusing, we designed a system that carefully synchronized the DMD, the photodetector, and the optimal pattern calculation, as illustrated in Fig. 3(a). In the current work, the whole process is based on a Hadamard algorithm modified from [29], where the Hadamard basis is a complete set representing binary amplitude patterns. The optimization scheme is mainly divided into two parts: 1) the characterization process and 2) parallel optimization.

In the characterization process, each Hadamard pattern is streamed through an HDMI cable to the DMD at a frame rate of 1,440 Hz, and the resulting signals are acquired with a photodetector. The total number of patterns is $2^{21}$ (= 2,097,152), which is twice the number of controlling modes $2^{20}$ (= 1,048,576). In order to handle the large amount of pattern data, the entire projection process was divided into 256 cycles, and pause times were inserted between successive cycles for pattern loading and data transfer. Then, the sequence of the measured signals was transferred piecewise to another personal computer (PC) via File Transfer Protocol (FTP), and used to calculate the optimal incident pattern in parallel. The effective frame rate was calculated to be 460 Hz, which is the total number of projected patterns over the total optimization time. The working principles of the DMD projections are presented in greater detail in Section 4.2.

In the parallel optimization process, the optimization algorithm involves a vector-matrix multiplication between the post-processed signal vector **S** with a dimension of 1 x $2^{20}$, and the Hadamard matrix **H** with a dimension of $2^{20}$ x $2^{20}$. Since the matrix size is too huge to be computed at once, the whole vector-matrix multiplication is also decomposed into a sub-problem $\mathbf{S}_i \mathbf{H}_i$ for each cycle $C_i$, and the calculations of each sub-problem are processed in another PC, simultaneously with the characterization process. The detailed algorithm is explained in Section 4.3.

## 4.1 Time-efficient optimization system

Time-efficient optimization in the proposed system is the product of three major components: (1) the rapid projection with a DMD, (2) high-speed acquisition with a single channel photodetector, and (3) the parallel calculation of the optimization pattern.

(1) The binary pattern projection speed (1440 Hz) of the DMD is much faster than the modulation speed (60 Hz) of the LCoS SLM. In practice, we also found that the speed of a LCoS SLM is slower than 60 Hz, due to the latency of the SLM controller.

(2) The single-channel photodetector offers a higher acquisition speed than the widely used CCD camera. When using the 2-D imaging array, it is hard to achieve both a fast acquisition rate and a fast data transfer rate. In practice, the effective frame rate of the CCD is usually slower than 100 Hz. The single photon counting module employed in the current study provides an acquisition speed of up to 10 MHz. Thus, the only limiting factor in the optimization time is the DMD projection speed.

(3) Using parallel computation to obtain wavefront optimization patterns can significantly reduce the overall optimization time. Current optimization algorithms essentially measure the transmission matrix (TM) of a scattering medium. Several of the TM based optimization algorithms need to complete the measurement of TM before optimization [24, 25]. In this study, the TM measurement and wavefront optimization were performed simultaneously. If the TM was fully measured and then the wavefront optimization was performed sequentially, the total experiment time would have been doubled, because the TM measurement time in a single cycle and the wavefront optimization time are almost the same in this study.

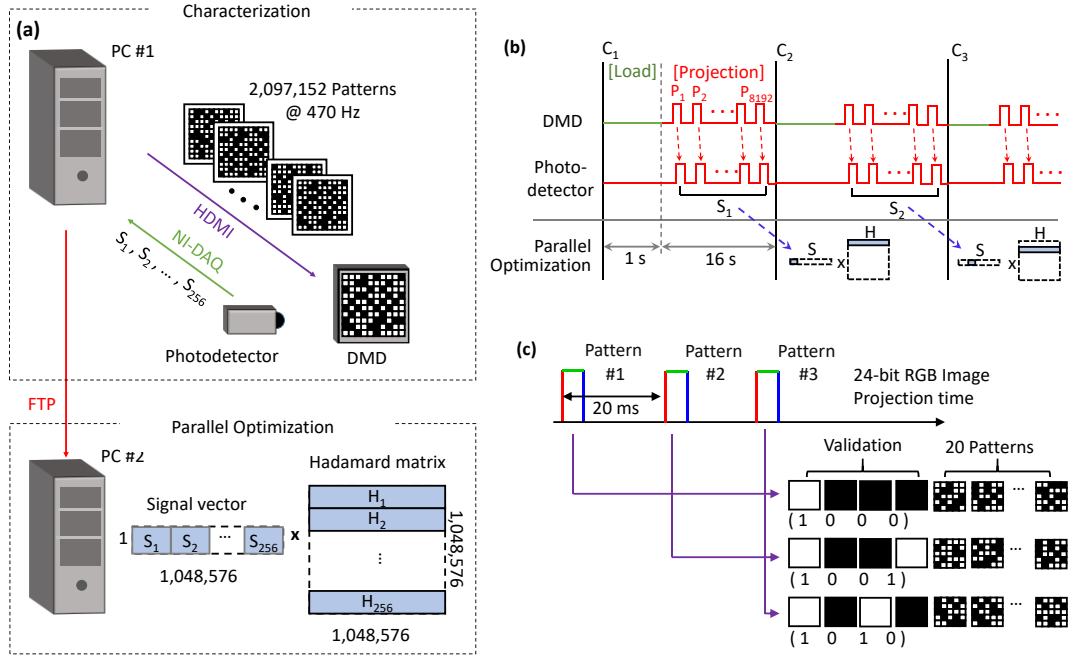

Fig. 3. (a) Configuration of the optimization system for ultrahigh enhancement of focusing. The DMD, a photodetector, and parallel optimization are synchronized using two PCs. PC # 1, for pattern projection and measurement, and PC # 2, for TM calculation, are operated at the same time. (b) Data flow diagram. Hadamard patterns ($P_1, P_2, …, P_{8192}$) are loaded into the PC memory and then streamed to the DMD. The sub-optimal pattern $S_i \times H_i$ is calculated in PC # 2 in parallel during the acquisition of a consequent signal $S_{i+1}$. (c) DMD projects at 1440 Hz by converting a 24-bit image input of 60 Hz to 24 binary level images. The first four patterns are used for error correction during synchronization, and the remaining 20 patterns are set to Hadamard-based patterns.

The workflow diagram of the whole system is shown in Fig. 3 (b). To perform one characterization cycle $C_i$, it takes 1 second to load patterns, and 16 seconds to project the 8,192 Hadamard patterns using the DMD. After the pattern projection, the measured signal is transmitted asynchronously to another computer (PC#2) as a vector $\mathbf{S}_i$ via FTP without interfering with the optimization time. While the next signal vector $\mathbf{S}_{i+1}$ is being measured, the calculation of the partial optimization pattern $\mathbf{S}_i\mathbf{H}_i$ is performed in PC#2. The average computation time is 16 seconds, which is slightly shorter than the single characterization period of 17 seconds. Thus, the total optimization rate of the current system is limited by the DMD projection time. The total optimization consists of 256 cycles of the characterization, which takes about 73 minutes.

### 4.2 Pattern projection in a DMD

To project a predefined Hadamard pattern onto the DMD, the pattern data must be transferred from the PC to the DMD. There are two ways to project specific pattern data using a DMD. 1) The pattern data can be preloaded to the onboard memory of the DMD via USB connection before projection, or 2) patterns can be streamed simultaneously via standard video cable during projection.

The current system uses the second approach because it is more suitable for controlling a large number of optical modes in TMs which use a large number of projection patterns. If patterns are preload on the onboard memory of the DMD, the maximum speed of the DMD at 23 kHz can be fully exploited. However, the maximum number of patterns that can be loaded onto the DMD is limited by the amount of onboard memory. The maximum memory capacity of a commercial DMD system is 8 GB (V-9600, 0.96″ WUXGA, 1920×1200, Vialux, Germany), which corresponds to about 30,000 patterns at 1920×1080 resolution, and that is not enough for one mega pattern projection. On the other hand, the memory available for streaming is only limited by the hard disk volume in the PC. For this reason, in the current study streaming the pattern during projection was better than preloading the pattern.

In the pattern streaming mode, a 24-bit color pattern is transmitted at 60 Hz via an HDMI cable. Since the Hadamard pattern is a 1-bit image with a binary value of 0 or 1, the amount of information contained in a single 24-bit color image is the same as the amount of information in 24 binary patterns. Some commercial DMD systems (DLP series, Texas Instruments Inc., USA) provide a function for converting a single 24-bit pattern into 24 binary patterns at the chipset level, and in this case the binary pattern projection can achieve a speed of 60 Hz × 24 = 1,440 Hz. This is relatively slower than the maximum refresh rate of 23 kHz of a DMD, but much faster than the refresh rate of a LCoS SLM.

The detailed implementation of the binary projection mode is illustrated in Fig. 3(c). The 24-bit color images #1–3 are streamed through the HDMI cable, and each color image contains 24 binary images. In an ideal situation, the streaming rate should be set to 60Hz. If this requirement is not satisfied, pattern duplication or missing patterns can occur, which can result in the complete destruction of the wavefront optimization. From an engineering point of view, however, it is difficult to accurately project a streaming pattern at 60 Hz. The accurate 60 Hz timing control, or 16,667 ns time interval control requires timed loop programming with 1 ns precision. This precision can only be achieved in a real-time operating system, where the maximum accuracy of Windows OS is 1 ms. Unfortunately, we have found that most real-time operating systems provide only limited functionality for video projection. For example, the Labview real-time module is not compatible with a display resolution of 1920×1080.

To solve this problem, we used the synchronization error correction method on Windows systems. In this method, 24-bit color images are streamed at a 50 Hz frame rate (20 ms interval). For one 24-bit image, as shown by the purple arrow in Fig. 3 (c), the first 4 patterns are set to the verification pattern, and the remaining 20 patterns are set to the Hammered pattern for the TM measurement. The role of the four validation patterns is to correct for projection errors. Because HDMI connections always import image data from a PC at a precise 60 Hz according to the V-sync protocol, all patterns are projected at least once, and some patterns are projected in duplicate. However, after observing the first 4 patterns, duplicate patterns can be found and eliminated. Each validation pattern represents a completely white or black color and is converted to a binary number between 0 and 1. Using four binary numbers, the sequence of 24-bit patterns is indexed sequentially from 0 to 15. By checking this validation code in the measurement signal, missing or duplicate patterns can be correctly identified in the calculation step. In our experiments, we always obtained > 95% accuracy after calibration.

### 4.3 Optimization algorithm

Let us define the input field projected onto a scattering layer using a DMD. $\mathbf{x} = [e_1 \ e_2 \ \cdots \ e_N]^T$ is a 1-bit $N \times 1$ vector (i.e. $e_i$ is either 0 or 1), where $N$ is the number of controlled pixels. Then, the single output channel (or mode) of the optical field transmitting through the scattering layer is denoted by a scalar $y$, which is related to the input field via the following linear equation:

$$y = \mathbf{Tx} \qquad (1)$$

where $\mathbf{T} = [t_1 \ t_2 \ \cdots \ t_N]$ is the TM of the scattering layer, and $y$ is a complex valued $1 \times N$ vector. Then, the intensity of the output field can be represented as,

$$|y|^2 = |t_1 e_1 + t_2 e_2 + \cdots + t_N e_N|^2 \qquad (2)$$

In order to maximize the intensity of the transmitted light at a specific position, $e_n$ in Eq. (2) should be carefully selected under the following criteria [28],

$$e_n = \begin{cases} 1 & \text{Re}(t_n) \geq 0 \\ 0 & \text{Re}(t_n) < 0 \end{cases} \qquad (3)$$

Therefore, the whole optimization process is now simplified to finding the sign of every element of $\text{Re}(\mathbf{T})$. To achieve this goal, we used a modified version of the Hadamard-based optimization algorithm [29]. The Hadamard matrix of order $m$, $\mathbf{H}^{(m)}$ is a matrix with a size of $2^m \times 2$. Superscripts $(m)$ are omitted in the following discussions, unless needed for further clarification. Each column in the matrix corresponds to one Hadamard basis vector, where $N = 2^m$ is equal to the number of control modes required or DMD pixels. In the experiment, $m = 20$ is mainly used, which corresponds to $N = 1,048,576$. Since the total number of pixels in the DMD is larger than $N$, the remaining DMD pixels are not used.

Now, let $\sigma_n$ be the output field corresponding to the incident wavefront shaped by projecting each Hadamard basis $\mathbf{h}_n$ onto the DMD. Similar to Eq. (2), the linear relationship between $\sigma_n$ and $\mathbf{h}_n$ holds as,

$$[\sigma_1 \ \sigma_2 \ \cdots \ \sigma_N] = \mathbf{T}[\mathbf{h}_1 \ \mathbf{h}_2 \ \cdots \ \mathbf{h}_N]. \qquad (4)$$

Using the inverse property of a matrix, $\mathbf{HH}^T = N\mathbf{I}$, and the real-valued property of $\mathbf{H}$, $\text{Im}(\mathbf{H}) = 0$ of the Hadamard matrix, the $\text{Re}(\mathbf{T})$ can be calculated as follows:

$$\text{Re}(\mathbf{T}) = \mathbf{SH}^T = \mathbf{SH}, \qquad (5)$$

where $\mathbf{S} = \frac{1}{N} \text{Re}[\sigma_1 \ \sigma_2 \ \cdots \ \sigma_N]$ is defined as a signal vector. Therefore, measuring $\mathbf{S}$ using the Hadamard basis directly provides $\text{Re}(\mathbf{T})$. This provides a great practical benefit because the full field of view of a DMD can be exploited for each Hadamard basis, whereas the pixel-wise measurement of $\text{Re}(\mathbf{T})$ would result in a significant decrease in signal-to-noise ratio because a part of the DMD pixel is used for the signal beam and the rest is used for the reference beam.

Inspired by the method in Ref. [29], we generated two binary DMD patterns, $\mathbf{v}_n^\pm$ for each $\mathbf{h}_n$ by adding reference field information with the uniform intensity,

$$\mathbf{v}_n^\pm = \frac{1}{2}(\mathbf{h}_1 \pm \mathbf{h}_n). \qquad (6)$$

Here, the uniform reference field is the first Hadamard basis vector, $\mathbf{h}_1 = [1 \ 1 \ \cdots \ 1]^T$. In order to measure $\text{Re}(\sigma_n)$, the measurements of two intensity signals are required,

$$\begin{aligned} I_n^\pm &= |\mathbf{Tv}_n^\pm|^2 \\ &= \left|\frac{1}{2}\mathbf{T}(\mathbf{h}_1 \pm \mathbf{h}_n)\right|^2 = \frac{1}{4}|\sigma_1|^2 + \frac{1}{4}|\sigma_n|^2 \pm \frac{1}{2}\text{Re}(\sigma_1^* \sigma_n). \end{aligned} \qquad (7)$$

By setting the phase of $\sigma_1$ as a reference phase without losing generality,

$$\text{Re}(\sigma_n) = \frac{I_n^+ - I_n^-}{\sqrt{I_1}}, \qquad (8)$$

where $I_1 = |\mathbf{Tv}_1|^2 = |\sigma_1|^2$ is the intensity corresponding to the $\mathbf{h}_1$ (uniform) pattern. In summary, the optimal condition in Eq. (3) can be calculated via Eqs. (5) and (8). In the proposed method, the minimum number of DMD projections is $2N-1$.

### 4.4 Parallel optimization

In the previous section, we explained that the optimal pattern calculation for $m = 20$ requires a matrix-vector multiplication between a $1 \times 2^{20}$ (= 1,048,576) vector $\mathbf{S}$ and a $2^{20} \times 2^{20}$ matrix $\mathbf{H}^{(20)}$. The size of this uncompressed matrix $\mathbf{H}^{(20)}$ is 4 terabytes in the single-precision floating-point format, and is an enormous size that cannot be loaded into PC memory at one time. To avoid large memory consumption and to efficiently parallelize the calculations, we converted the whole matrix multiplication into the sum of the partial matrix multiplications as follows:

$$\mathbf{SH} = [\mathbf{S}_1 \ | \ \mathbf{S}_2 \ | \ \cdots \ | \ \mathbf{S}_{256}] \begin{bmatrix} \mathbf{H}_1 \\ \mathbf{H}_2 \\ \vdots \\ \mathbf{H}_{256} \end{bmatrix} = \mathbf{S}_1 \mathbf{H}_1 + \cdots + \mathbf{S}_{256} \mathbf{H}_{256}. \qquad (9)$$

The signal vector **S** is decomposed into 256 subvectors $\mathbf{S}_n$ with dimensions of $1 \times 4{,}096$. Similarly, the Hadamard matrix **H** is also divided into 256 submatrices $\mathbf{H}_n$ with dimensions of $4{,}096 \times 2^{20}$. Another advantage of the present partial matrix multiplication is that the entire matrix $\mathbf{H}^{(m)}$ is not necessarily generated. Even when using built-in functions in MATLAB or LabVIEW, the creation of the matrix $\mathbf{H}^{(m)}$ with $m = 20$ is problematic due to memory limitations. We found that the maximum order of the Hadamard basis that can be evaluated in MATLAB is $m = 17$ with 16 GB of memory. Therefore, by exploiting the following recursive relation, we separately computed $\mathbf{H}_1^{(20)}, \mathbf{H}_2^{(20)}, \cdots, \mathbf{H}_{256}^{(20)}$ from $\mathbf{H}^{(17)}$,

$$\mathbf{H}^{(m+1)} = \begin{bmatrix} \mathbf{H}^{(m)} & \mathbf{H}^{(m)} \\ \mathbf{H}^{(m)} & -\mathbf{H}^{(m)} \end{bmatrix}. \quad (10)$$

## 5. Experimental results

### 5.1 Measurements of the enhancement factor

The EF of the optical focus formed with the optimized wavefront is defined as the ratio between the intensity of the formed optical focus and the mean background intensity [1]. The mean background intensity was measured as the mean intensity for the projection of all Hadamard basis vectors. Since the optimized focus has extremely strong light intensity, ND filters (OD = 3.0; ND30A, Thorlabs, Inc., United States) were inserted between the laser and the single mode fiber. EFs greater than 800 cannot be measured without an ND filter because the dynamic range of the photodetector is 40 Mc/s (number of photons per second) and the average background intensity is 50 Kc/s. To obtain the precise value of EF, the optical density of the ND filter was calibrated for each optimization test. In this calibration, the optimal DMD pattern with $N = 2^{12}$ is projected, resulting in an EF of approximately 650. The corrected optical density is then obtained from the two intensity signals, which are obtained with and without the ND filter. The actual optical density was between 577 and 695.

### 5.1 Point optimization results

To visualize the optimized focus, we took photographs of the optical foci using a DSLR camera (EOS 1100D, Canon, Japan, 30 mm F1.4 EX DC, Sigma, USA), as shown in Figs. 4(a-c). As shown in Fig. 4 (a), the uncontrolled illumination beam produces a speckle pattern after the second diffuser. After the incident wavefront was optimized, a clear focus was formed at the desired position, as shown in Fig. 4(b). The number of optical modes used in the control was 1,048,576, and the experimentally measured EF was 10,531. The total optimization time was 73 minutes.

To demonstrate the capability of multi-focal formation, three optical foci were formed at a distance of 2 mm apart. After optimizing each optical focus individually at three different positions, a new wavefront, which results in the simultaneous formation of three optical foci, was calculated by coherently summing the three obtained wavefronts. As shown in Fig. 4(c), it was confirmed that three optical foci were formed at one time. The total optimization time to generate three foci was 3 hours and 45 minutes. All photographs were taken at the same exposure time. Since the EF of the optical focus formed at a single point decreases in inverse proportion to the total number of optimized foci [1], when three optical foci are formed, the intensity of light of each focus is smaller than when a single focus is optimized.

It is worth noting that we imaged the foci with a DSLR camera, which is a de-magnifying optical system. In previous studies, the EF was too small and consequently required a high magnification microscopy system to image the enhanced focus. However, the ultrahigh enhancement achieved in this study enables direct imaging of the optical foci without extra magnification. We found that the optical foci can even be seen with unassisted eyes.

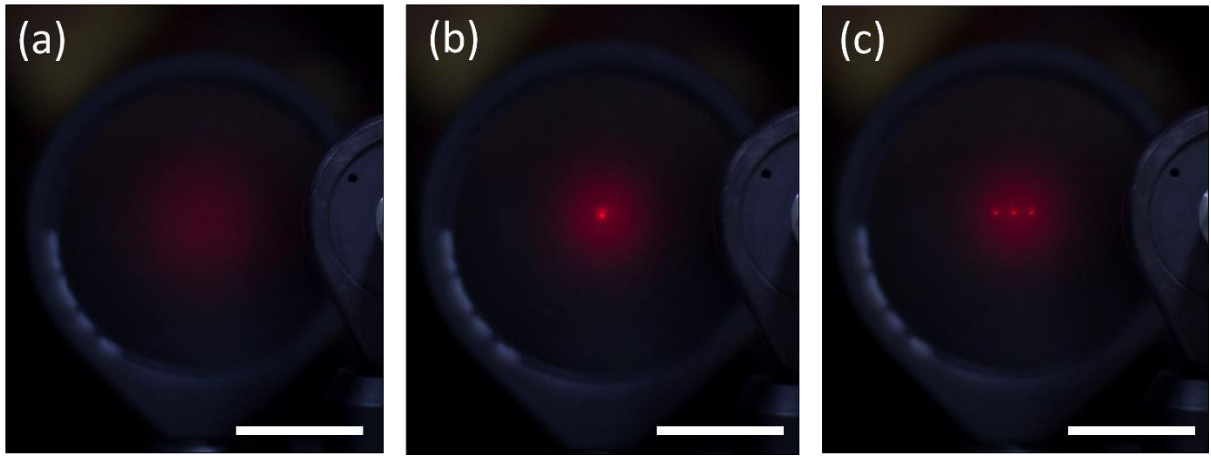

Fig. 4. (a) A photograph of the speckle patterns with uncontrolled illumination. (b-c) Photographs of a single optimized focus (b) and three optimized foci (c). The scale bar is 2 cm.

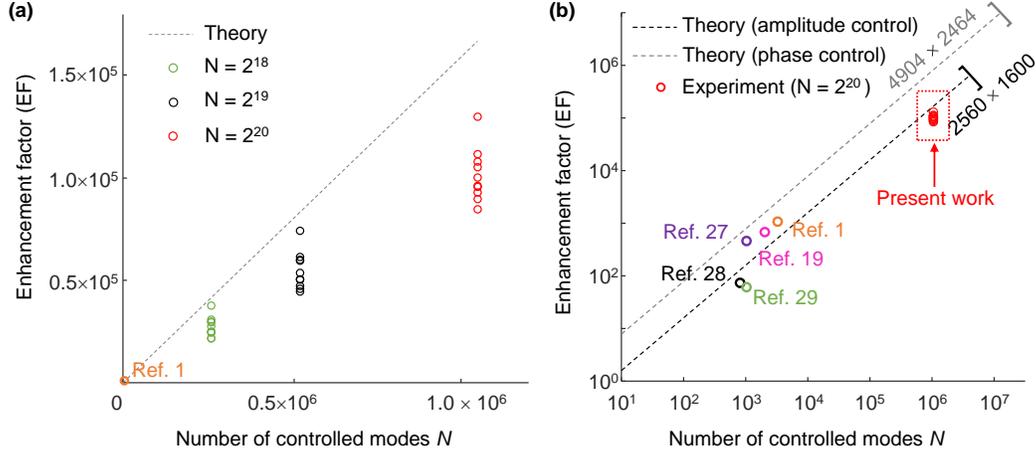

Fig. 5. (a) Intensity enhancement as a function of the number of controlling modes. The green, black and red color solid circles denote the experimental EFs for 10 trials for $N = 2^{18}$, $2^{19}$ and $2^{20}$ modes, respectively. The blue dashed line represents the theoretical prediction. (b) The EFs obtained in the present study are indicated by the solid red circles. EFs found in the literature [1, 19, 27-29] are represented by other colored circles.

6. Quantitative analysis

In order to test the performance of the proposed technique, experiments for achieving optical focus after turbid media were performed for various control modes. The results are shown in Fig. 5. The theoretical EF is given as $\eta = 1 + (N/2 - 1)/\pi$ for the binary amplitude modulation [28], which is indicated by the blue dotted line in Fig. 5. The experimental EF values for the three experiments corresponding to $N = 2^{18}$, $2^{19}$ and $2^{20}$ were 27,322 ± 4,834; 54,812 ± 9,222; 101,391 ± 12,940, respectively.

In particular, the experiment with EF = 101,391, which corresponds to $N = 2^{20}$, is about 100 times greater than the previous best record, where EF = 1,080 for $N = 3,228$ (the orange circle in Fig. 5). In our experiment, the resulting EF values correspond to 65.49% ($N = 2^{18}$), 65.69% ($N = 2^{19}$) and 60.75% ($N = 2^{20}$) of the theoretical predictions, respectively. These results clearly indicate that the proposed system provides stable optimization even for an extremely large number of control modes.

To compare the capability of the present system with existing wavefront shaping techniques, we plotted the EFs in Fig. 5(b) in logarithmic scale. The theoretical EF for the phase control and the binary amplitude control are represented with green and blue dashed lines, respectively. The EFs achieved with phase control [1, 19, 27] and amplitude control [28, 29] all ranged from 10–1000, which are all far smaller than the values in the present work. To our best knowledge, the current maximum number of pixels among commercially available SLMs is 4904 x 2464 (GAEA, HoloEye Photonics AG, Germany) for LCoS SLMs and 2560 x 1600 (DLP9000, Texas Instruments Inc., United States) for DMDs. Achieving an EF of 101,391 with $2^{20}$ = 1,048,576 modes in this study is about 100 times better than the existing best and is close to the maximum performance that can be achieved with existing wavefront shapers.

**7. Discussion**

In this study, we developed an efficient optical system for optimizing focus through scattering media using the Mega-pixel modes of a high-speed DMD. The number of pixels used ($2^{20}$) approached the limit of the single commercially available DMD, and the EF that was achieved was 100 times better than the existing best reported EFs. The optimization results showed robust performance for various control mode numbers, and EF reached approximately 60% of the theoretical predictions.

The high EF values demonstrated in this study can be compared with the digital optical phase conjugation (DOPC) approach. The DOPC system records multiply-scattered light originating from a point source and plays back its phase conjugate field, which converges into the initial focus. The EF of DOPC is linearly proportional to the number of pixels, as in feedback-based focus optimization.

Careful calibration is required to generate clear focus in the DOPC systems because accurate alignment between the CCD field measurement and the SLM field projection is critical [35, 36]. Our system does not require complex calibration because optical aberrations and severe multiple scattering are compensated automatically by the optimization algorithm. The maximum EF in our study is comparable to the state-of-art result of 120,000 exhibited by the DOPC system [35]. Furthermore, the proposed approach can also be implemented in phase conjugation experiments [14].

A possible further improvement in the proposed system would be to increase the speed of optimization. Using a real-time OS that can provide an accurate 60 Hz projection rate is expected to result in a slight improvement in optimization speed. In this case, the total optimization time would be about 30 minutes for the experiments using the effective frame rate of 1.2kHz and $N = 2^{20}$. In principle, under ideal circumstances with a DMD full frame rate of 23 kHz, the overall experiment time can be reduced to 45 seconds when using 1,000,000 modes. Advances in data transfer technology with custom FPGA programming can significantly improve the projection speed.

Because the projection time is comparable to the computation time at the current stage, the computation time and the DMD projection time must be shortened simultaneously to improve the overall experiment speed. Reduced computation time can be achieved through parallel computing using multiple PCs or through fast computing systems such as the Amazon Web Service. We also found that introducing a graphics processing unit (GPU) was not an effective solution to increase speed, because it takes a very long time to load a significantly large matrix into the GPU memory, so the fast computation time of the GPU cannot be exploited.

Increasing the number of controllable pixels is another direction for future work to improve the system performance. When controlling a larger number of pixels, it is important to manipulate large pattern data and calculate a large number of vector matrix multiplications. In the current work, the size of the total patterned data streamed to the DMD was 414 GB, which was stored on a solid-state drive (SSD) for fast pattern loading. If the number of optical modes is doubled, the size of the pattern data is quadrupled to 1.6 terabytes. The size of the Hadamard matrix **H** in parallel optimization also increases from 4 terabytes to 16 terabytes, with an increase in computational cost. We expect this large amount of data can be managed using appropriate engineering up to $N = 2^{22}$. However, achieving an even greater number of pixels would be technically very difficult. Therefore, developing a new optimization algorithm that bypasses the limited dynamic range and signal-to-noise ratio of the detector without relying on TM measurements is another possible strategy.

The ultrahigh enhancement factors proposed and experimentally demonstrated in this study can provide promising tools for various applications of wavefront shaping technology. Using the proposed method, a high-quality image can be projected through a scattering medium [5, 30]. The energy transfer efficiency in the local area can be greatly enhanced by focus optimization without requiring the complex wavefront shape of a perfect transmission channel [6, 37]. We also envision that this technology to be applied to a variety of interesting areas, such as near-field control, optical lithography, and photodynamic therapy.

**Funding**. This work was supported by KAIST, and the National Research Foundation of Korea (2015R1A3A2066550, 2014M3C1A3052567, 2014K1A3A1A09063027).